\documentclass[showpacs,prl,amsmath,amssymb,aps,floatfix]{revtex4}
\usepackage{graphicx}
\usepackage{dcolumn}
\usepackage[sort&compress]{natbib}
\usepackage[usenames,dvipsnames]{xcolor}
\usepackage{bm}
\usepackage{color}
\usepackage[dvips]{epsfig}
\newcommand{\comment}[1]{}
\newcommand{\AAA}{{\rm \,\AA}}
\definecolor{brightpink}{rgb}{1.0, 0.0, 0.5}

\begin{document}

\title{Can morphological changes of erythrocytes be driven by
  hemoglobin?}


\author{S. G. Gevorkian$^{1,2}$, A.E. Allahverdyan$^{2}$,
  D.S. Gevorgyan$^{3}$, Wen-Jong Ma$^{1,4}$, Chin-Kun
  Hu$^{1}$ } 

\affiliation{$^{1}$Institute of Physics, Academia Sinica, Nankang,
Taipei 11529, Taiwan}

\affiliation{$^{2}$Yerevan Physics
Institute, Alikhanian Brothers St. 2, Yerevan 375036, Armenia}

\affiliation{$^{3}$ Institute of Fine Organic Chemistry, 26
Azatutian ave., Yerevan 0014, Armenia}

\affiliation{$^{4}$ Graduate Institute of Applied Physics, National
  Chengchi University, Taipei 11605, Taiwan}

\date{\today}

\begin{abstract} 

  At 49$^\circ$ C erythrocytes undergo morphological changes due to an
  internal force, but the origin of the force that drives changes is
  not clear. Here we point out that our recent experiments on
  thermally induced force-release in hemoglobin can provide an
  explanation for the morphological changes of erythrocytes.

\end{abstract}


\maketitle

It is well-known since 1865 that at 49$^\circ$ C erythrocytes undergo
morphological changes (vesiculation and deformation) due to an
internal force. This effect is employed in medicine, but the origin of
the force that drives morphological changes is not clear. Here we
point out that our recent experiments on thermally induced
force-release in hemoglobin can provide an explanation for the
morphological changes of erythrocytes.

The oxygen transport in our organisms is carried out by
hemoglobin \cite{eaton}. It consists of four globular units
linked into a double-dimer tetrameric structure \cite{eaton}; see
Fig.~1. Each unit can carry one oxygen molecule
${\rm O_2}$ attached to its heme group. The oxygen binding ability is
cooperative \cite{eaton}. It decreases upon reducing the pH
factor or increasing the concentration of ${\rm CO}_2$
\cite{bohr}. Due to this Bohr's effect \cite{bohr} a tissue with a
stronger need of oxygen receives it more.

Hemoglobin is densely packed in erythrocyte. In contrast to other
cells, erythrocytes do not have a nucleus for the purpose of greater
storage. The orientation of hemoglobin molecules in erythrocytes is
not random \cite{fok,fok2}.

Erythrocyte is known to change its physical features after thermal
treatment at 49$^\circ$--50$^\circ$ C. This was discovered in 1865 via detecting a
rich spectrum of erythrocyte morphological changes at and above 49$^\circ$--50$^\circ$
C \cite{schultze}. The effect is routinely employed for studying the
spleen enlargement, because when thermally treated and radioactively
tagged erythrocytes are immersed back to blood, they are trapped in
the spleen. This trapping was prescribed both to changing the form of
erythrocyte (from disc to sphere) \cite{harris} and to plasticity loss
\cite{romania}. Morphological changes at 49$^\circ$--50$^\circ$ C were studied by
scanning electron micrography in \cite{coakley}. It was found that
thermal effects at 49$^\circ$--50$^\circ$ C depends rather weakly on heating rate
(provided that this rate is sufficiently slow, i.e. slower than 0.75 C
per second) and that morphological changes proceed via two major
scenarios. Either the biconcave erythrocyte form changes to a rosette
shape with well-established protuberances, or the erythrocyte
fragments into several parts \cite{coakley}. Nearly 50 \% of
erythrocytes did not undergo any visible morphological change at
49$^\circ$--50$^\circ$ C \cite{coakley}. The effect of morphological changes was
prescribed to denaturation of spectrin, a cytoskeletal protein that
stitches the intracellular side of the plasma membrane in eukaryotic
cells including erythrocyte \cite{spectrin,coakley2,ivanov}. The
effect can be suppressed by lowering the ionic strength, by presence
of albumin \cite{coakley2}, or (to a large extent) after incorporation
of adamantin derivatives into cell membranes \cite{herrmann}. However,
the detailed mechanism of the morphological change is not yet
understood; in particular this concerns the physical part of the
problem, i.e. the origin of the force that drives vesiculation and
deformation.

We carried out micromechanical experiments on crystals of horse and
human hemoglobin \cite{hem}. These experiments show that precisely at
49$^\circ$ C the hemoglobin releases force \cite{hem}. The main
advantage of using biopolymer crystals is that there is a possibility
of displaying those motions of the macromolecule that can have only
transient character in the solution
\cite{eaton,perutz3}.
These motion are controlled by the water content and intermolecular
contacts, which in their turn are regulated by the crystal syngony.
Thus the solid state hemoglobin is close to its {\it in vivo} state in
mammal erythrocytes, where the hemoglobin is densely packed with
concentration $\simeq 34\%$ \cite{trincher}.

In its partially unfolded state|i.e. for a temperature higher than the
physiological temperatures, but lower than the unfolding
temperature|the hemoglobin responds to heating by a sudden release of
force and a subsequent jump of the Young's modulus \cite{hem}. The
detailed structure of this effect is different for human and horse
hemoglobin, but the temperature where the effect takes place is equal
to 49$^\circ$ C for both types of hemoglobin \cite{hem}. This temperature does
not depend on the hydration level (in contrast to denaturation
temperatures) and also on the solvating level. We argued that the
effect relates to certain slowly relaxing (mechanical) degrees of
freedom of the quaternary structure of hemoglobin that accumulate
energy during heating and then suddenly release it at 49$^\circ$ C
\cite{hem}. Surprisingly, a force-release effect was found under
heating which is generally supposed to diminish mechanical features of
biopolymers. It was already noted in \cite{swed,yan,artmann}
that 49$^\circ$ C may indicate on structural changes in hemoglobin, but the
important aspect of the force release was noted only in \cite{hem}.
Such an effect is absent in the thermal response of myoglobin.
Myoglobin also binds and unbinds oxygen, but does so without a sizable
cooperativity. This relates to its function: myoglobin is a depot (not
transporter) of oxygen in muscles.

\begin{figure}
  \centerline{\includegraphics[width=.7\textwidth]{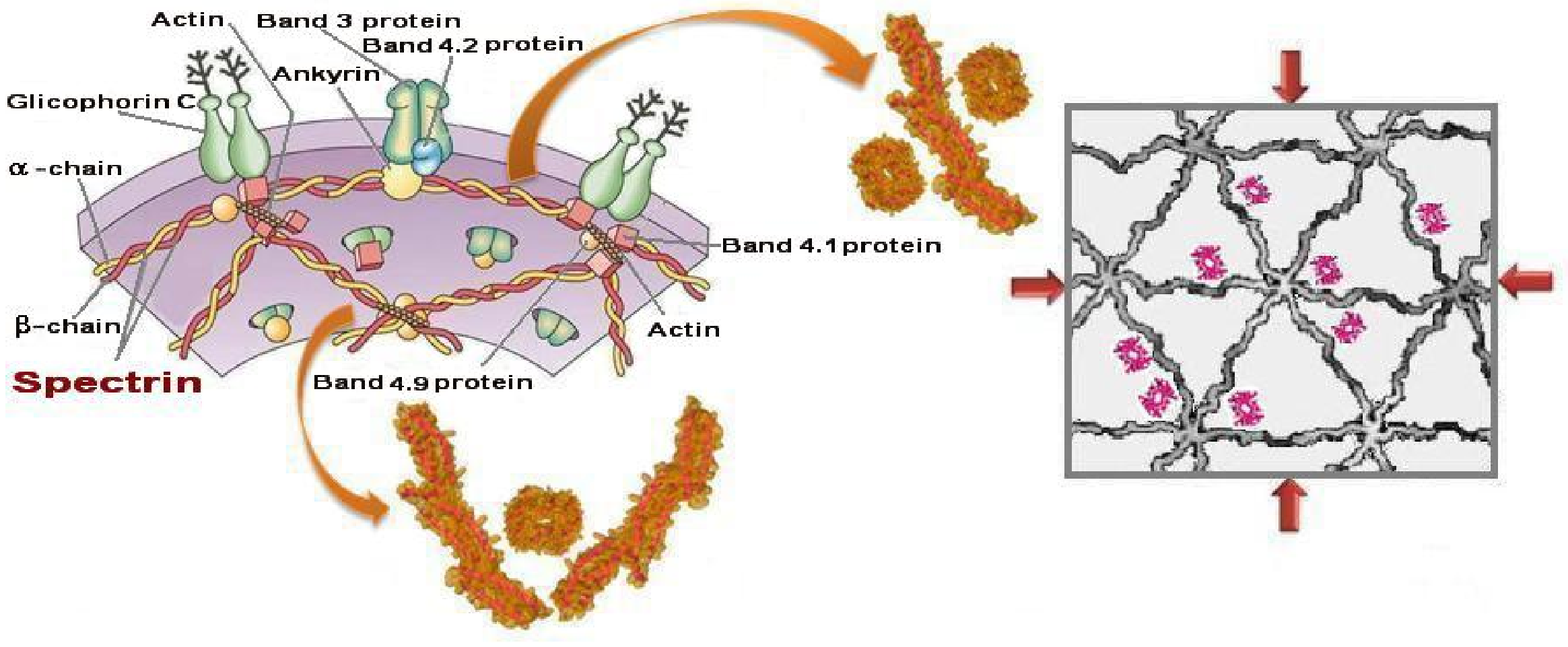}}
  \vskip 2 mm \caption{(Color online) Schematic representation of the
    hemoglobin-spectrin interaction.\\ Left figure: Multiprotein
    complexes in the red cell membrane that are attached to
    spectrin. Tetrameric band 3 protein is bound to ankyrin, which
    binds the quasi-hexagonal spectrin network to erythrocyte
    membrane. Spectrin consists of α and β chains. It stitches the
    membrane, which is very soft without spectrin. The hemoglobin is
    depicted with partially unfolded quaternary structure: two out of
    four globular sub-units are partially open. \\ Right figure:
    possible location of hemoglobin molecules within the spectrin
    network. In order not to overload the picture, only few hemoglobin
    molecules are shown. In reality the hemoglobin concentration is
    approximately 34 \%. Red arrows show the force that can stitch the
    erythrocyte.  }
\label{fig}
\end{figure}

Here we conjecture that the driving force for the morphological
transitions of erythrocytes at 49$^\circ$-50$^\circ$ C does come from
hemoglobin. The spectrin denaturation still does play a role in those
morphological changes, e.g. because it transfers the force over the
erythrocyte membrane; see Fig.~\ref{fig} for a schematic
representation. Spectrin denaturation {\it alone} cannot explain
morphological changes, since (normally) denaturation does not relate
to force release. Also, the spectrin is found in membranes of other
cells that do not carry hemoglobin (e.g. neurons
\cite{neuron_spectrin}), but no high-temperature morphological effects
are known for them.

We cite the following arguments in support of this conjecture:

-- The onset of morphological changes is at 49$^\circ$ C, and precisely at
this same temperature the hemoglobin releases force. The temperature
does not depend on the type of hemoglobin (this was checked for human
and horse hemoglobins), and it also does not depend on hydration and
solvation levels. 

-- Before the force release, the internal friction of hemoglobin shows
a sizable increase \cite{hem}. Hence the quaternary structure of
hemoglobin {\it partially} denaturates, and it is prone to forming
spectrin-hemoglobin complexes.  Spectrin-hemoglobin complexes were
studied by various means \cite{hmspectrin2,hmspectrin1,hmspectrin0}.
Their life-time can vary widely \cite{hmspectrin1,hmspectrin0}. It is
also known that spectrin-hemoglobin complexes do not seriously alter
the hemoglobin oxygenation.

If this conjecture is confirmed by further experiments, it may mean
that the interaction between hemoglobin and spectrin is relevant for
oxygen carrying function of erythrocytes. We note that dependencies
of erythrocyte membrane features on the hemoglobin content were
suggested several times, but no experimental support was so far found
on them \cite{contra1,contra2}. However, these experiments did not
study the thermal features at 49$^\circ$ C, which is the main focus of
this contribution.

Once our conjecture looks for the origin of a mechanic force, it can
be relevant for mechanic explanation of several important aspects of
hemoglobin, including hemoglobin binding on spectrin \cite{24},
scenarios of pathological intracellular polymerization in the process
of hemoglobin deoxygenation \cite{25}, and non-equilibrium dynamics of
hemoglobin \cite{27}. For the first case our conjecture can explain
how the influence is transferred from the hemoglobin to
erythrocyte. In the second case it may prevent the intracellular
polymerization. For the third case, the force may be involved in
generating the non-equilibrium potential.

\section*{Acknowledgements}
This work was supported in part by Grant MOST 105-2112-M-001 -004.

\section*{Author contributions}

SGS designed research, performed research, analyzed data.

AEA analyzed data, wrote the paper.

DSG performed research, analyzed data.

WJM analyzed data, wrote the paper.

CKH analyzed data, wrote the paper.



\clearpage

\end{document}